\documentclass[twocolumn,showpacs,showkeys,prc,aps,amsmath,amssymb,10pt]{revtex4-1}
\usepackage{graphicx}
\usepackage{color}
\usepackage{multirow}
\usepackage{longtable}
\usepackage{soul}
\begin{document}

\title{Shell model predictions for $^{124}$Sn double-beta decay}

\author{Mihai Horoi}
\email{mihai.horoi@cmich.edu}
\author{Andrei Neacsu}
\email{neacs1a@cmich.edu}
\affiliation{Department of Physics, Central Michigan University, Mount Pleasant, Michigan 48859, USA}
\date{\today}

\begin{abstract}
Neutrinoless double-beta ($0 \nu \beta \beta$) decay is a promising beyond Standard Model  process. Two-neutrino double-beta ($2 \nu \beta \beta$) decay is an associated process that is allowed by the Standard Model, and it was observed in about 10 isotopes, including decays to the excited states of the daughter.  $^{124}$Sn was the first isotope whose double-beta decay modes were investigated experimentally, and despite few other recent efforts, no signal has been seen so far. Shell model calculations were able to make reliable predictions for $2 \nu \beta \beta$ decay half-lives. Here we use shell model calculations to predict the $2 \nu \beta \beta$ decay half-life of $^{124}$Sn. Our results are quite different from the existing quasiparticle random-phase approximation (QRPA) results, and we envision that they will be useful for guiding future experiments. We also present shell model nuclear matrix elements for two potentially competing mechanisms to the  $0 \nu \beta \beta$ decay of $^{124}$Sn.
 
\end{abstract}

\maketitle

\section{Introduction}
Neutrinoless double-beta decay is a low-energy nuclear process that could identify beyond Standard Model (BSM) physics, especially the lepton number violation (LNV) 
effects and exotic neutrino properties, such as the neutrino mass scale and whether neutrino is a Dirac or a Majorana 
fermion \cite{SchechterValle1982}. 
Neutrino oscillation experiments have successfully measured the squared mass differences among neutrino mass eigenstates \cite{Kaether2010, SuperKamiokande2011, DoubleChooz2012, DayaBay2012, SNO2010, KamLAND2011},
yet the nature of the neutrinos and the absolute neutrino masses cannot be obtained from such measurements. 
This has led to both theoretical and experimental efforts dedicated to the discovery of the $0 \nu \beta \beta$ decay mode, as reflected by the large numbers of recent 
reviews  \cite{Avignone2008, Ejiri2010, Vergados2012}. 
The simplest and most studied mechanism is the exchange of light Majorana neutrinos in the presence of left handed 
weak interaction, but other possible mechanisms contributing to the total $0 \nu \beta \beta$ decay rate are taken into consideration \cite{Barry2013,Horoi2013,Stefanik2015,HoroiNeacsu2015}. 
Such  mechanisms include the contribution of the right-handed currents \cite{Mohapatra1975, Doi1983}, and mechanisms involving super-symmetry 
\cite{Vergados2012, Hirsch1996}. 
$0 \nu \beta \beta$ decay and the analysis of the same-sign dilepton decay channels at hadron colliders 
\cite{Cms2011, Atlas2011, LHCbMinus2012, LHCbPlus2012} are the best approaches to investigate these mechanisms. However, $0 \nu \beta \beta$ decay results from several 
isotopes would be necessary to disentangle contributions of different mechanisms \cite{Faessler2011, HoroiNeacsu2015, Stefanik2015}. 
Additional information regarding the neutrino physics parameters could be obtained 
from large-baseline and new reactor neutrino oscillation experiments \cite{Sno2013}, and from cosmology \cite{Planck2013}. 

Two-neutrino double-beta decay is an associated process that is allowed by the Standard Model, and it was observed in about 10 isotopes, including decays to the excited states of the daughter \cite{Barabash2015}.  
$^{124}$Sn was the first isotope whose double-beta decay modes were investigated experimentally \cite{Fireman1948,Fireman1952}. The Q-value for the decay to the ground state (g.s.), 2.289 MeV,  is slightly smaller than those of $^{130}$Te and $^{136}$Xe, but it is larger than that of $^{76}$Ge decay. As a consequence, the phase space factor (see below) for $2\nu\beta\beta$ decay of $^{124}$Sn is about ten times larger than that of $^{76}$Ge, while the $0\nu\beta\beta$ phase space factor of $^{124}$Sn is about four times larger than that of $^{76}$Ge. 
Double-beta decay modes of $^{124}$Sn were also recently investigated experimentally by several groups \cite{Barabash2008, Dawson2008, DawsonDegering2008, Hwang2009}, 
but no signal has been seen so far. The best upper limit obtained was about $1\times10^{21}$ yr \cite{Barabash2008}, very close to the QRPA prediction of $2.7\times10^{21}$ yr for 
the $2 \nu \beta \beta$ transition to the first excited $0^+_1$ state of the daughter \cite{Aunola1996}, which   was used as guidance (see also Table 21 in Ref. \cite{SuhonenCivitarese1998}). The QRPA prediction \cite{Aunola1996,SuhonenCivitarese1998} for the $2 \nu \beta \beta$ transition to the g.s. of the daughter was available at the time of these experiments,  $T^{2\nu}_{1/2}=7.8\times10^{19}$ yr, but this transition was also not observed. 

$^{124}$Sn is still considered for the experimental investigations of its double-beta decay modes by the TIN.TIN experiment \cite{Vandana2014}, and reliable predictions of $2\nu\beta\beta$ decay half-lives and $0\nu\beta\beta$ nuclear matrix elements (NME) would energize the experimental effort and could save millions of dollars and time.
Shell model calculations were able to make a reliable prediction for $2 \nu \beta \beta$ decay half-life of $^{48}$Ca \cite{Caurier1990}, which was later confirmed by the experimental data \cite{Balysh1996}. We calculated the $2 \nu \beta \beta$ for several isotopes of immediate experimental interest, including $^{48}$Ca \cite{HoroiStoicaBrown2007,Horoi2013}, $^{76}$Ge \cite{SenkovHoroi2014}, $^{82}$Se \cite{SenkovHoroiBrown2014}, $^{130}$Te, and $^{136}$Xe \cite{NeacsuHoroi2015}. Here we use shell model calculations to predict the $2 \nu \beta \beta$ decay half-lives of $^{124}$Sn, including transitions to the g.s., and to the first excited $2^+_1$ and $0^+_1$ states of the daughter. Our results are different from those  of the existing QRPA \cite{Aunola1996} and shell model \cite{CaurierNowacki1999} calculations.

Recent experimental proposals to investigate the $0\nu\beta\beta$ decay of $^{124}$Sn \cite{Vandana2014} raised the need for accurate calculations of the 
$0 \nu \beta \beta$ nuclear matrix elements for this isotope to guide the experimental effort. 
The accuracy of the calculated nuclear matrix elements still represents the largest uncertainty in the theoretical analyses of the neutrinoless double-beta decay.  
These NME are currently investigated by several methods, 
such as Interacting Shell Model (ISM) \cite{Retamosa1995, Caurier2008, MenendezPovesCaurier2009, Caurier2005, HoroiStoica2010, Horoi2013, SenkovHoroi2013, HoroiBrown2013, SenkovHoroiBrown2014, NeacsuStoica2014, SenkovHoroi2014, NeacsuHoroi2015}, 
QRPA \cite{Simkovic1999, Suhonen2010, Faessler2011, MustonendEngel2013, FaesslerGonzales2014}, 
Interacting Boson Model (IBM-2) \cite{Barea2009, Barea2012, Barea2013, Barea2015}, Projected Hartree Fock Bogoliubov (PHFB) \cite{Rath2013}, 
Energy Density Functional (EDF) \cite{Rodriguez2010}, and the Relativistic 
Energy Density Functional (REDF) \cite{Song2014} method. 
There are still large differences among the NME calculated with different methods and by different groups, and that has been a topic of many debates in the literature 
(see e.g. \cite{Faessler2012,Vogel2012}). 

We calculate the NME for $2 \nu \beta \beta$ and $0 \nu \beta \beta$ decay of $^{124}$Sn in a shell model approach, using a recently proposed effective Hamiltonian, which was fine-tuned with experimental data for Sn isotopes.
Realistic effective nucleon-nucleon ($nn$) interactions, derived from free $nn$ potentials, form the microscopic basis of shell model calculations \cite{Jensen1995}. 
However, these effective Hamiltonians often require additional fine-tuning to the available data to gain real predictive power.
For the calculations reported in this paper, we use a recently proposed shell model effective Hamiltonian (called SVD here) \cite{Chong2012} and the shell model space $jj55$ consisting of the 
%for nucleons between the $Z,N = 50,\ldots,82 $ shell closures with 
$0g_{7/2}, 1d_{5/2}, 1d_{3/2}, 2s_{1/2}$ and $0h_{11/2}$ valence orbitals.
To ensure the reliability of the results, we investigate this Hamiltonian by performing calculations of spectroscopic quantities, and 
comparing them to the latest experimental data available for the nuclei involved in the decay, i.e. $^{124}$Sn and $^{124}$Te. 
These tests include the energy spectra, the $B(E2)\uparrow$ transition probabilities, the Gamow-Teller ($GT$) strengths, and the occupation probabilities for both neutrons and protons \cite{NeacsuHoroi2015}. 

The paper is organized as follows. In the following Section we briefly present the formalism for NME involved in the expressions of $\beta \beta$ decay half-lives 
via exchange of both light Majorana neutrinos and heavy neutrinos mechanisms. 
Section \ref{detail} presents some shell model results of the SVD effective Hamiltonian validating the spectroscopy of $^{124}$Sn and $^{124}$Te.  
The NME are shown in Section \ref{nme-sn},  $2\nu\beta\beta$ calculations in subsection \ref{2nme}, and the $0\nu\beta\beta$ results in subsection \ref{0nme}, including an overview of recently reported NME results in \ref{overview}.  
Section \ref{conclusions} is dedicated to conclusions.

\section{$\beta \beta$ decay formalism}

The $0\nu\beta\beta$ half-lives are usually expressed as a product of a leptonic phase space factor (PSF), a NME that depends on the nuclear structure of 
the mother and that of the daughter nuclei, and a LNV parameter related to the BSM mechanism considered. To obtain reliable limits for the LNV parameters, 
accurate measurements of the $0\nu\beta\beta$ decay half-lives accompanied by precise calculations of the PSF and NME are needed. 
Considering the exchange of light left-handed neutrinos and heavy right-handed neutrinos, the following expression for $0 \nu \beta \beta$ 
decay half-life is a good approximation \cite{Horoi2013}:
\begin{equation}
\label{0nhl}
\left[ T^{0\nu}_{1/2} \right]^{-1}=G^{0\nu}_{01} g_A^4\left( \left| M^{0\nu}_\nu\right|^2 \left| \eta_{\nu L} \right|^2 + \left| M^{0\nu}_N\right|^2 \left| \eta_{N R} \right|^2\right) .
\end{equation}
Here $G^{0\nu}_{01}$ is the phase space factor for this decay mode \cite{Kotila2012, StoicaMirea2013,SuhonenCivitarese1998} that depends on the decay energy  and nuclear charge, $M^{0\nu}_{\nu,N}$ are the the NME, $\eta_{\nu L}$ and $\eta_{N R}$ are the 
neutrino physics parameters associated to the light neutrino-exchange and the heavy neutrino-exchange mechanisms, respectively \cite{Faessler2011,Horoi2013}.
The expressions for $M^{0\nu}_{\nu,N}$ have the following structure:
\begin{equation}
\label{0nnme}
 M^{0\nu}_{\nu,N}=M^{0 \nu}_{GT}-\left( \frac{g_V}{g_A} \right)^2 \cdot M^{0 \nu}_F + M^{0 \nu}_T \ ,
\end{equation}
where $g_V$ and $g_A$ are the vector and the axial-vector coupling strengths, respectively, while $M^{0 \nu}_{GT}$, $M^{0 \nu}_F$ and  $M^{0 \nu}_T$ are the Gamow-Teller ($GT$), the Fermi($F$) and the Tensor ($T$) components, respectively.
Explicit expressions for $M_\alpha^{0\nu}$ ($\alpha=GT,\ F,\ T$) can be found in several papers, for example Ref. \cite{SenkovHoroiBrown2014}. They are defined to be dimensionless.
%In our calculations, we have employed all the nuclear structure ingredients and parameters used in the recent literature \cite{sim-09}. 
Our method includes short-range correlations (SRC), finite nucleon size effects (FNS), and higher order corrections of the nucleon current 
\cite{HoroiStoica2010, Horoi2013, SenkovHoroi2013, HoroiBrown2013, SenkovHoroiBrown2014, NeacsuStoica2014}.

 \begin{figure}[ht!]
 \includegraphics[width=0.95\linewidth]{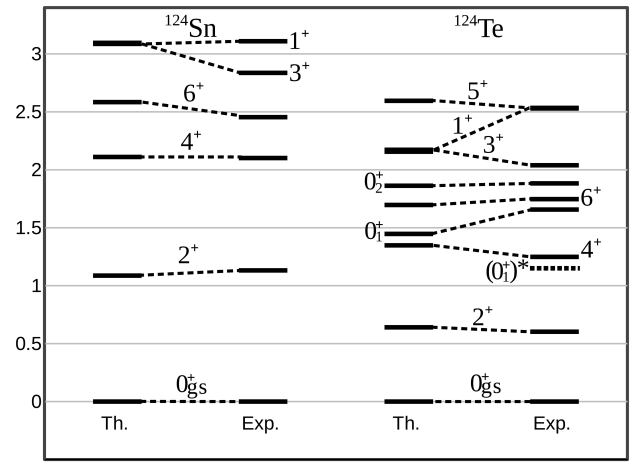}
 \caption{The calculated energy levels (in MeV) for $^{124}$Sn and $^{124}$Te (left columns) compared to experiment (right columns).}
 \label{fig-elevels}
 \end{figure}

For $2 \nu \beta \beta$
decay, the half-life for the transition to a state of angular momentum $J$ ($J=0\ \text{or } 2$) of the daughter nucleus is given to a good approximation by \cite{HoroiStoicaBrown2007}
\begin{equation}
\label{2nhl}
\left[ T^{2\nu}_{1/2} \right]^{-1} = G^{2\nu}(J) g_A^4 \left| (m_ec^2)^{J+1} M_{2\nu}(J) \right|^2 ,
\end{equation}
where $G^{2\nu}(J)$ is a phase space factor \cite{Doi1985,SuhonenCivitarese1998,Mirea2014up}, and $M_{2\nu}(J)$ is the $2\nu\beta\beta$ NME, which can be calculated as \cite{SuhonenCivitarese1998,HoroiStoicaBrown2007}
\begin{equation}
 M_{2\nu}(J) = \frac{1}{\sqrt{J+1}} \sum_k \frac{\left< J_f || \sigma \tau^- || 1^+_k\right> \left< 1^+_k || \sigma \tau^- || 0^+_i \right>}{\left( E_k + E_J \right)^{J+1} }  .
\label{2nnme}
\end{equation}
Here the $k$-sum is taken over the $1^+_k$ states with excitation energies $E_k$ in the intermediate nucleus, $^{124}$Sb in this case. $E_J = \frac{1}{2}Q_{\beta\beta}(J)+\Delta M$, where $Q_{\beta\beta}(J)$ is the Q-value for the transition to the state of angular momentum $J$ in the daughter nucleus, and $\Delta M$ is the difference in mass between the intermediate nucleus and the decaying nucleus. The nuclear matrix elements can be calculated efficiently using a strength function approach \cite{Horoi2011pa}.

\section{Analysis and validation of the effective Hamiltonian}\label{detail}
For the shell model calculations of the $^{124}$Sn $\beta\beta$ decay NME, we need a suitable effective Hamiltonian, one which can reliably describe the structure of the nuclei involved in the decay. 
One option is the use an effective Hamiltonian in a large model space that includes the $0g_{9/2}, 0g_{7/2}, 1d_{5/2}, 1d_{3/2}, 2s_{1/2}, 0h_{11/2}$, and $0h_{9/2}$ orbitals (called $jj77$). 
This approach was successfully used in the case of $^{136}$Xe \cite{HoroiBrown2013}. However, the shell-model dimensions for $^{124}$Sn in this model space are too large.
%for realistic shell model calculations.
Another option is using the $jj55$ model space and an effective Hamiltonian fine-tuned to the experimental data. 
Extending our previous analysis from Ref. \cite{NeacsuHoroi2015}, we now investigate the SVD Hamiltonian reported in Ref. \cite{Chong2012}. This effective
Hamiltonian was obtained starting with a realistic CD-Bonn $nn$ potential \cite{Machleidt2001}, and the core-polarization effects 
have been taken into account via many-body perturbation theory \cite{Jensen1995}.  The resulting $jj55$ Hamiltonian was further fine-tuned using the experimental data for Sn isotopes. 
 Ref. \cite{Chong2012} presents a detailed study of this effective Hamiltonian for Sn isotopes. Here we analyze how accurately it describes the nuclei of interest for this study by comparing to the available experimental data.

\begin{figure} [ht!]
 \includegraphics[width=0.95\linewidth]{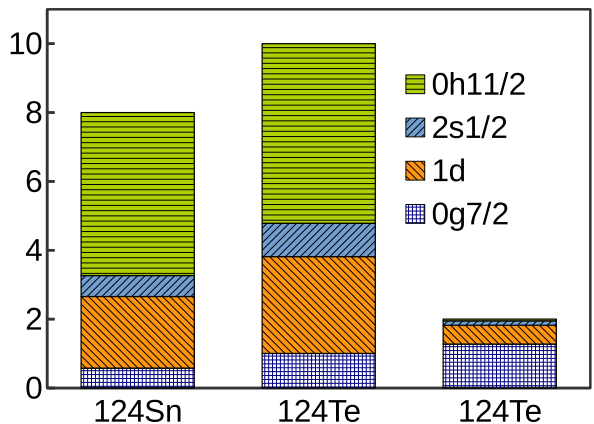}
 \caption{(Color online) Theoretical neutron shell vacancies for $^{124}$Sn and $^{124}$Te, and proton shell occupancies for $^{124}$Te (right column).}
 \label{fig-occup}
\end{figure}

Using this effective Hamiltonian, we calculate and compare with the experimental data, when available, the following spectroscopic quantities for the nuclei in the region of interest: 
the energy spectra for some $\left[ 0^+ - 6^+ \right]$ states, $B(E2)\uparrow$ transition probabilities, occupation probabilities and the Gamow-Teller strengths. 
For some of these quantities, such as the occupancies and the GT strengths, experimental data is not yet available, but we present our theoretical results hoping that future experimental investigations will be validating these predictions.
This Hamiltonian was previously investigated in a similar analysis \cite{NeacsuHoroi2015} done for $^{130}$Te and $^{136}$Xe.

 \begin{figure} [ht!]
 \includegraphics[width=0.95\linewidth]{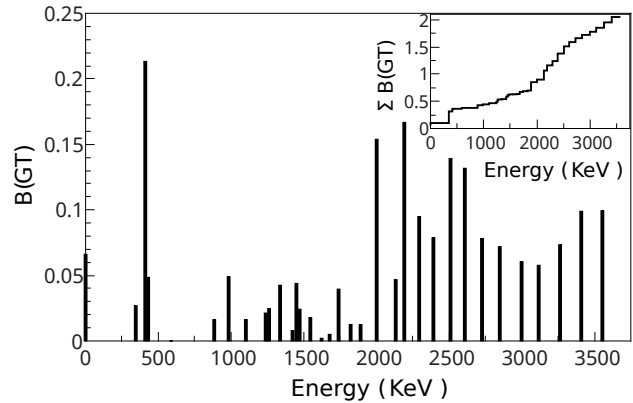}
 \caption{Calculated $^{124}$Sn GT strengths. The inset presents the calculated GT running sum.}
 \label{fig-gt}
 \end{figure}

\subsection{Energy levels} \label{elevels}

We use the SVD Hamiltonian to investigate the low-energy states for the initial and the final nuclei, $^{124}$Sn and $^{124}$Te, respectively.
Fig. \ref{fig-elevels} presents the low-energy spectra calculated with the SVD Hamiltonian compared the experimental data \cite{Katakura2008}. 
In the case of the initial nucleus,$^{124}$Sn, our shell model calculations describe the experimental levels with very good accuracy. This is expected 
since this Hamiltonian was fine-tuned using experimental energies of yrast states for Sn isotopes, including $^{124}$Sn. 

Describing the daughter nucleus, $^{124}$Te, is more challenging. First, because it was not included in the fine-tuning, and in addition, the experimental data available is not very clear about the first excited $0^+$ state, which is very important for our analysis. In particular, although the recent double-beta decay experimental studies \cite{Barabash2008,DawsonDegering2008,Dawson2008} assume that this state is at 1.156 MeV excitation energy, the latest Nuclear Data Sheets \cite{Iimura1997,Katakura2008} analyzing this mass region exclude this level and put the first excited $0^+$ state at 1.657 MeV.   Refs. \cite{Barabash2008,DawsonDegering2008,Dawson2008} relied on two editions of Table of Isotopes \cite{Tableofisotopes7,Tableofisotopes8} when considering this state, but the associated gamma rays included in the Table of Isotopes are extremely weak and unreliable. 
Moreover, QRPA theoretical predictions of the $2\nu\beta\beta$ transitions to the first excited $0^+$ state of $^{124}$Te \cite{Aunola1996,SuhonenCivitarese1998,Suhonen2011} were using until recently the Table of Isotopes assignment.
Our calculation puts this level at 1.44 MeV, in-between 1.156 MeV and 1.657 MeV. Given the known uncertainty of the shell model results, one cannot exclude the 1.156 MeV assignment. The calculated second excited $0^+$ state, however, is just very close to the experimental energy listed in the Nuclear Data Sheets (see Fig. \ref{fig-elevels}). This result favors the Nuclear Data Sheets assignment. However, we decided to still consider the Table of Isotopes assignment for energy of the first excited $0^+$ state as an alternative possibility. This state, the first excited calculated $0^+$ state assumed to have 1.156 MeV excitation energy when calculating the phase space factors, is denoted with $\left(0^+_1\right)^*$ throughout this paper (see Fig. \ref{fig-elevels}, and Tables \ref{tab_2nme} and \ref{tab_0nme}).

For the yrast states of $^{124}$Te we get reasonable results, 
only the first $1^+$ state being underestimated by 350 MeV. In addition, for the second $2^+$ (not included in Fig. \ref{fig-elevels}) we get an excitation energy of 1.294 MeV, very close to the experimental value of 1.326 MeV. This leads us to consider this effective Hamiltonian as well suited for the description of the
spectra for these initial and final nuclei involved in the $\beta\beta$ decay.
\begin{table}[ht!]
 \caption{The calculated $B(E2)\uparrow \ (e^2 b^2)$ values  of $^{124}$Sn and $^{124}$Te nuclei compared to the adopted values \cite{Pritychenko2013, Pritychenko2014}.}
 \begin{tabular}{l|cccc} \hline \hline
    & $^{124}$Sn  & $^{124}$Te & $^{126}$Te & $^{128}$Te \\
\hline
           $B(E2)\uparrow_{th.}$  &  0.146  & 0.579 & 0.505 & 0.340 \\
$B(E2)\uparrow_{ad.}$  & 0.162  & 0.560 & 0.474 & 0.380 \\ \hline \hline
\end{tabular}
 \label{tab_be2}
 \end{table}

\subsection{$B(E2)\uparrow$ transitions} \label{be2}

When calculating the $B(E2)\uparrow$ values, we use the neutron effective charge ($e^{eff}_n=0.88e$) recommended \cite{Chong2012} for Sn isotopes when using this 
effective Hamiltonian. For the proton effective charge, we consider $e^{eff}_p=1.88e$, which appears to describe well the $B(E2)\uparrow$ transitions in 
Te isotopes. The results for $^{124}$Sn, $^{124}$Te, $^{126}$Te, and $^{128}$Te are displayed in Table \ref{tab_be2} where we find very good agreement with the adopted $B(E2)\uparrow$ data \cite{Pritychenko2013,Pritychenko2014}.

\subsection{Occupation probabilities} \label{occupancies}
A good description of the experimental occupation probabilities, although not always sufficient, it is generally necessary to validate shell model predictions.
For the nuclei of interest we could not find reliable experimental data for the comparison of the neutron vacancies in $^{124}$Sn and $^{124}$Te, 
and of the proton occupancies in $^{124}$Te. 
We include our theoretical predictions for the occupation probabilities in Table \ref{tab-occup} and in Fig. \ref{fig-occup}.
%in the analysis of the Hamiltonian performed in this section alongside the study the other spectroscopic observables. 
The occupation probabilities of the $1d_{5/2}$ and $1d_{3/2}$ orbitals are summed up and presented as $1d$, similar to the 
Figures 2-4 of Ref. \cite{NeacsuHoroi2015}.
\begin{table}[h!]
  \caption{Theoretical neutron ($n$) shell vacancies for $^{124}$Sn and $^{124}$Te, and proton ($p$) shell occupancies for $^{124}$Te 
  denoted with $p$.}
\begin{tabular}{lcccc}\hline\hline
	 & $0g_{7/2}$	& $1d$	& $2s_{1/2}$	& $0h_{11/2}$ \\ \hline
  $n$ $^{124}$Sn&0.578	&2.078	&0.609	&4.735 \\
  $n$ $^{124}$Te&1.003	&2.813	&0.970	&5.214 \\
  $p$ $^{124}$Te&1.276	&0.545	&0.128	&0.050 \\ \hline \hline
 \end{tabular}
  \label{tab-occup}
\end{table}

\subsection{$GT$ strengths} \label{gt}
The validation of the Gamow-Teller strength distribution is particularly relevant for a good description of double-beta decay rates. 
Using this effective Hamiltonian, the spin-orbit partners orbitals of $0g_{9/2}$ and $0h_{9/2}$ are missing in the $jj55$ model space, and therefore the Ikeda sum-rule
is not satisfied. In a previous paper \cite{NeacsuHoroi2015}, we have studied the effects of GT sum-rule  for $^{130}$Te and $^{136}$Xe, and we were able to compare with  the experimental data.
In this case of the $^{124}$Sn nucleus no reliable experimental data could be found, and we can not present a comparison.
Fig. \ref{fig-gt} shows our calculated GT strengths for the transition of $^{124}$Sn to $^{124}$Sb. 
%The running GT sum is displayed in the inset of the plot.
%We present in Fig. \ref{fig-gt} our calculated GT strengths for the transition of $^{124}$Sn to $^{124}$Sb. 
The running GT sum is displayed in the inset of the plot. A quenching factor $q_f=0.74$ was used in the calculations presented in Fig. \ref{fig-gt}, the same value that
was utilized in the $2\nu\beta\beta$ NME calculations from Section \ref{2nme}.

 \begin{figure} [ht!]
 \includegraphics[width=0.95\linewidth]{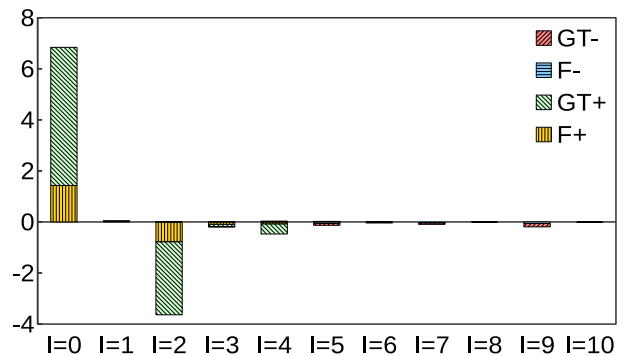}
 \caption{(Color online) $I$-pair NME decomposition: Gamow-Teller and Fermi (multiplied by $-(g_V/g_A)^2$) matrix elements for the $0\nu\beta\beta$ decay of $^{124}$Sn (light neutrino-exchange) for configurations
with pairs of neutrons/protons coupled to some spin ($I$) and some parity (positive or negative). CD-Bonn SRC parameterization was used.}
 \label{pair_light}
 \end{figure}

 \begin{figure} [ht!]
 \includegraphics[width=0.95\linewidth]{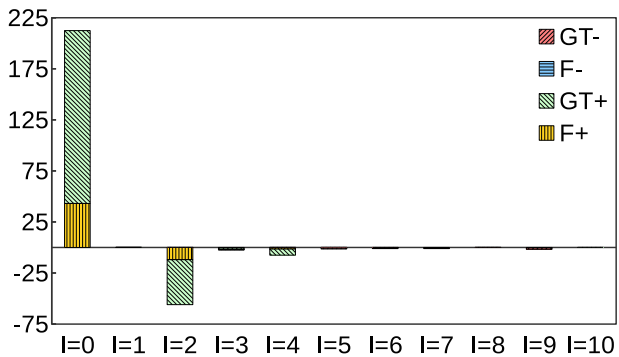}
 \caption{(Color online) Same as Fig. \ref{pair_light}, for heavy neutrino-exchange.}
 \label{pair_heavy}
 \end{figure}

\section{Nuclear matrix elements for $^{124}\text{Sn}$} \label{nme-sn}

Following the investigation of the SVD Hamiltonian for spectroscopic quantities of the initial $^{124}$Sn and the final $^{124}$Te nuclei, we present the 
NME for the double-beta decay of $^{124}$Sn. Calculations of the $2\nu\beta\beta$ decay are presented in Subsection \ref{2nme}. 
We consider these results as particularly important for the upcoming Sn $\beta\beta$ experiments \cite{Vandana2014}, as predictions of the half-lives can be made.
Subsection \ref{0nme} presents the results of our calculations for the $0\nu\beta\beta$ decay NME.
These calculations are compared with other results from the literature in Table \ref{tab_nme_comp_light} for the light neutrino-exchange mechanism, and in Table \ref{tab_nme_comp_heavy} for the heavy neutrino-exchange case. For a better overview of the spread of the NME obtained with different nuclear structure methods and by different groups, we also display these values in Fig. \ref{light}  (light) and Fig. \ref{heavy}  
(heavy). $g_A=1.27$ was used in Eqs. (\ref{0nhl}) and (\ref{2nhl}) to calculate all the half-lives reported below.

\subsection{$2\nu\beta\beta$ decay of $^{124}$Sn NME}\label{2nme}

As we mentioned in the Introduction, double-beta decay modes of $^{124}$Sn were recently investigated experimentally by several groups \cite{Barabash2008, Dawson2008, DawsonDegering2008, Hwang2009}. The emphasis was on decays to the excited states,
but no signal has been seen so far. The best upper limit obtained for the half-life,  $1\times10^{21}$ yr, was reported in Ref. \cite{Barabash2008} (which seems that was using the QRPA prediction for 
the $2 \nu \beta \beta$ transition to the first excited $0^+_1$ state of the daughter given in Refs. \cite{Aunola1996,SuhonenCivitarese1998}, $2.7\times10^{21}$ yr, as guidance).  A QRPA prediction for the $2 \nu \beta \beta$ half-life for the transition to the ground state of the $^{124}$Te was available at the time of these experiments, $7.8\times10^{19}$ yr, but this transition was also not observed. However, as discussed in section \ref{elevels}, all these studied, experimental and theoretical, used an assignment for the first excited $0^+_1$ state that was not validated by the recent Nuclear Data Sheets \cite{Katakura2008}.
We also mentioned in the Introduction, that the shell model calculations were able to make reliable predictions for $2 \nu \beta \beta$ decay half-life of $^{48}$Ca \cite{Caurier1990} before the experimental measurements. At the time  when the above experiments for $^{124}$Sn were performed a shell model half-life for the $2\nu\beta\beta$ decay to the g.s. of $^{124}$Te was available in Ref.\cite{CaurierNowacki1999}, but this conference proceedings reference provided little information about the calculations.

We calculated the $2 \nu \beta \beta$ NME for several isotopes of immediate experimental interest, including $^{48}$Ca \cite{HoroiStoicaBrown2007,Horoi2013}, $^{76}$Ge \cite{SenkovHoroi2014}, $^{82}$Se \cite{SenkovHoroiBrown2014}, $^{130}$Te, and $^{136}$Xe \cite{NeacsuHoroi2015}. Here we use shell model calculations to predict the $2 \nu \beta \beta$ decay half-lives of $^{124}$Sn, including transitions to the g.s.,  and to the first excited $2^+_1$ and $0^+_1$ states of $^{124}$Te. One of the main sources of uncertainty of the associated NME is the quenching factor $q_f$ of the Gamow-Teller operator, $\sigma \tau^-$ in Eq. (\ref{2nnme}), which was observed in many shell model calculations. The typical $q_f$ value for the $pf$-shell model space is 0.74 \cite{Cole2012}, but it varies in model spaces such as $jj55$ due to the missing spin-orbit partners that breaks the Ikea sum-rule \cite{HoroiBrown2013}. However, for a fine-tuned Hamiltonian, the Gamow-Teller strength gets distorted towards low-energy, and the $q_f$ that describes the $2\nu\beta\beta$ NME seems to be more stable and closer to 0.7 \cite{SenkovHoroiBrown2014,SenkovHoroi2014,NeacsuHoroi2015,BrownFangHoroi2015}. In Ref. \cite{NeacsuHoroi2015} we found for the same SVD Hamiltonian and the same $jj55$ model space that the quenching factors that describe well the $2 \nu \beta \beta$ NME of $^{130}$Te and $^{136}$Xe were 0.71 and 0.68, respectively. In Ref. \cite{NeacsuHoroi2015} we compared the calculated NME for $^{130}$Te with that reported in Ref. \cite{Barabash2010} (for $^{136}$Xe we use the NME from  Ref. [64] in \cite{NeacsuHoroi2015}) to extract these quenching factors. An updated analysis of the experimental data \cite{Barabash2015}, which is using updated half-lives and phase space factors \cite{Mirea2014up} for both isotopes, became recently available. We reanalyzed the calculations of $^{130}$Te and $^{136}$Xe NME \cite{NeacsuHoroi2015} using the newly recommended values, and we found that the quenching factors need to be slightly changed to 0.79 and 0.69, respectively. Given the new range of the quenching factors for $^{130}$Te and $^{136}$Xe we chose $q_f=0.74$ for the calculations of the $2 \nu \beta \beta$ NME and half-lives of $^{124}$Sn. Based on this range of $q_f$ values one can estimate a theoretical uncertainty of at least $\pm 20\%$ for the $2\nu\beta\beta$ half-lives of $^{124}$Sn. A more microscopic theory of the shell model quenching factors in heavy nuclei is needed, but not yet available. A recent paper \cite{MenendezGazit2011} suggests that the quenching could be related to the two-body currents induced by the three-body forces, but there is not wide agreement with that interpretation, and it is known that the distortion of the Gamow-Teller strength in calculated in reduced valence spaces could play a significant role \cite{BrownFangHoroi2015}.

Table \ref{tab_2nme}  presents our predictions for the $2\nu\beta\beta$ half-lives of $^{124}$Sn. Transitions to ground state and to the first excited $0^+_1$ and $2^+_1$ states are considered. Table \ref{tab_2nme} also shows  for comparison the old  \cite{SuhonenCivitarese1998} and the updated \cite{Suhonen2011} QRPA half-lives, which we associated with the $\left(0_1^+\right)^*$ (see discussion in section \ref{elevels}),  as well as shell model \cite{CaurierNowacki1999} and isospin-fixed QRPA  \cite{SimkovicRodin2013} half-lives for the transition to the g.s. of $^{124}$Te. Included in the table are also the state-dependent Q-values ($Q_{\beta\beta}$), and the associated phase space factors ($G^{2n}$) and NME. Updated $2\nu\beta\beta$ PSF for the excited states of $^{124}$Sn are only available for the first excited $0^+$ state identified by Ref. \cite{Katakura2008} (see Table II of Ref. \cite{Kotila2012}). Therefore we used the more comprehensive results of Ref. \cite{Mirea2014up} for guidance to obtain the phase space factor for the  $\left(0_1^+\right)^*$  state. The PSF for the transition to the first excited $2+$ state in Ref. \cite{Mirea2014up} are inaccurate, but we obtained updated values \cite{StoicaPrivate2015,MireaPahomi2015}. We were able to reproduce with about 3\% accuracy the PSF from Tables I, II, and III of Refs. \cite{Mirea2014up,MireaPahomi2015} using Eqs. (A.1)-(A.5) and (A.27)-(A.28) of Ref. \cite{SuhonenCivitarese1998}, and a 0.95 screening factor for the charge of the daughter \cite{HoroiNeacsuPSF}. The PSF for $^{124}$Sn obtained with this method were used in Table \ref{tab_2nme}. The $2\nu\beta\beta$ transition to the $2^+$ state is suppressed due to the larger power entering the denominator of the NME, Eq. (\ref{2nnme}). A discussion of the suppression of the transitions to $2^+$ states can be found in Refs. \cite{Doi1983,Doi1985}. The $2\nu\beta\beta$ transition to the first excited $0_1^+$ state in $^{124}$Te is also highly suppressed due to small Q-value resulting in a very small phase space factor. If the first excited $\left(0_1^+\right)^*$ exists at the excitation energy suggested by the Table of Isotopes, then the predicted half-life for this state could be within experimental reach. 

\begin{table*}[htb]
 \caption{The $2\nu\beta\beta$ half-lives (in units of $10^{21}\ yr$) for decays to the g.s., the first $0^+_1$ excited state, for the alternate first excited $\left(0_1^+\right)^*$ state, and the first $2^+_1$ excited state of $^{124}$Te, compared with other results reported in the literature. Ref. \cite{Suhonen2011} provides two values, WS/Adj, which are listed here in this order.}
%\begin{ruledtabular}
\begin{tabular}{ccccc|c|c|c|c} \hline\hline
$J^{\pi}_i$	  & $Q_{\beta\beta}$(MeV) & $G^{2\nu}\ (10^{-21}\ yr^{-1})$ & $M^{2\nu}$ (MeV$^{-(J+1)}$) & $T_{1/2}^{2\nu}$ & $T_{1/2}^{2\nu}$ \cite{SuhonenCivitarese1998} & $T_{1/2}^{2\nu}$ \cite{Suhonen2011} & $T_{1/2}^{2\nu}$ \cite{SimkovicRodin2013} & $T_{1/2}^{2\nu}$ \cite{CaurierNowacki1999} \\ \hline 
$0^+_{g.s.}$ & 2.289   & 531   & 0.0423  & 1.6   & 0.078 & 0.13/0.043  & 0.18 & 0.29 \\ 
$0_1^+$      & 0.630   & 0.0199  & 0.0418	 & $6.2\times 10^5$   &  &   & & \\
$\left(0_1^+\right)^*$      & 1.120   & 1.51  & 0.0418	 & 558   & 2.7 & 88/94  & & \\
$2_1^+$	     & 1.686   & 9.12  & 0.00328 & $2.2\times 10^5$  & $6.5\times 10^5$ & $(4.8/44)\times 10^2$   &  & \\ \hline  \hline
\end{tabular}
\label{tab_2nme}
\end{table*}

\begin{table*}[htb]
 \caption{The {\em on-axis} $0\nu\beta\beta$ half-lives in $yr$ for light ($T_{1/2}^{0\nu}$) and heavy ($T_{1/2}^{0N}$) neutrino-exchange decays to the g.s., and the first $0^+_1$ excited state of $^{124}$Te, using the {\em on-axis} neutrino physics parameters extracted from the $^{136}$Xe half-life: $\mid \eta_{\nu L} \mid =10^{-6}$ and  $\mid \eta_{N R} \mid =1.2\times 10^{-8}$. These estimates are based on NME calculated with the CD-Bonn SRC.}
%\begin{ruledtabular}
\begin{tabular}{ccccccc} \hline\hline
$J^{\pi}_i$	  & $Q_{\beta\beta}$(MeV) & $G^{0\nu}\ (10^{-15}\ yr^{-1})$ & $\ \ \ M^{0\nu}_{\nu}\ \ \ $ & $\ \ \ M^{0\nu}_{N}\ \ \ $ & $T_{1/2}^{0\nu}$ & $T_{1/2}^{0N}$    \\ \hline 
$0^+_{g.s.}$ & 2.289   & 9.06   & 2.17  & 144   & $2.3\times 10^{25}$ & $3.7\times 10^{25}$ \\ 
$0_1^+$      & 0.630   & 0.216  & 0.456	 & 32.6 & $2.2\times 10^{28}$   &  $3.0\times 10^{28}$ \\
$\left(0_1^+\right)^*$      & 0.63   & 0.953  & 0.456	 & 32.6   & $5.0\times 10^{27}$ & $6.9\times 10^{27}$ \\
 \hline  \hline
\end{tabular}
\label{tab_0nme}
\end{table*}

\begin{table*}[htb]
  \caption{$0\nu\beta\beta$  NME obtained with different nuclear structure methods for the light neutrino-exchange mechanism. When two values are given the lowest corresponds to Argonne-V18 SRC and the other to CD-Bonn SRC. See text for details.}
\begin{ruledtabular}
 \begin{tabular}{lcccccccc} 
	&$^{48}$Ca	&$^{76}$Ge	&$^{82}$Se	&$^{124}$Sn	&$^{130}$Te	&$^{136}$Xe	\\ \hline 
ISM-CMU &0.80/0.88	&3.37/3.57	&3.19/3.39	&2.00/2.15	&1.79/1.93	&1.63/1.76	\\
ISM-StMa&0.85		&2.81		&2.64		&2.62		&2.65		&2.19	\\
QRPA-Tu &0.54/0.59	&5.16/5.56	&4.64/5.02	&2.56/2.91	&3.89/4.37	&2.18/2.46	\\
QRPA-Jy &		&5.26		&3.73		&5.30		&4.00		&2.91	\\
IBM-2   &1.75		&4.68		&3.73		&3.19		&3.70		&3.05		\\	
%QRPA-UNC&		&5.09/5.53	&		&		&1.37/1.38 &1.55/1.68	\\     
 \end{tabular}
 \end{ruledtabular} 
 \label{tab_nme_comp_light}
 \end{table*}

\begin{table*}[htb]
 \caption{Same as Table \ref{tab_nme_comp_light} for the heavy neutrino-exchange mechanism.}
  \begin{ruledtabular}
 \begin{tabular}{@{}l@{}cccccccc@{}} 
	       &$^{48}$Ca	&$^{76}$Ge	&$^{82}$Se	&$^{124}$Sn	&$^{130}$Te	&$^{136}$Xe	\\ \hline 
ISM-CMU \  &52.9/75.5	&126/202	&127/187	&97.4/141	&94.5/136	&98.8/143	\\
ISM-StMa \  &56.5	&132.7		&122.4		&141		&144.2		&114.9		\\
QRPA-Tu &40.3/66.3	&287/433	&262/394	&184/279	&264/400	&152/228	\\
QRPA-Jy &		&401.3		&287.1		&453.4		&338.3		&186.3	\\ 
IBM-2   &46.3/76.0	&107/163	&84.4/132	&79.6/120	&92.0/138	&72.8/109		\\  
 
 \end{tabular}
  \end{ruledtabular} 
 \label{tab_nme_comp_heavy}
 \end{table*}

\subsection{$0\nu\beta\beta$ decay NME for $^{124}$Sn}\label{0nme}

The $0\nu\beta\beta$ decay NME calculations performed in this paper are done within the closure approach. 
We use a recently proposed method \cite{SenkovHoroi2014} to obtain the optimal closure energy \cite{SenkovHoroiBrown2014}, $\left<E\right>$, for $^{124}$Sn by calculating 
the optimal closure energy for $^{136}$Xe with this effective Hamiltonian, and we find $\left<E\right> = 3.5 \ MeV$.

%Unlike the $2\nu\beta\beta$ decay shell model predictions, in the $0\nu\beta\beta$ decay calculations it is yet unclear if NME need to be quenched.

The analysis of the $0\nu\beta\beta$ NME is extended, similar to that of Ref. \cite{NeacsuHoroi2015}, 
by looking to the decomposition of the NME over the angular momentum $I$ of the proton (or neutron) pairs (see Eq. (B4) 
in Ref. \cite{SenkovHoroi2013}), called $I$-pair decomposition. In this case, the NME can be written as $M_\alpha=\sum_{I} M_\alpha (I)$, where $M_\alpha (I)$ 
represent the contributions from each pair-spin $I$ to the $\alpha$ part of the NME. 
Fig. \ref{pair_light}, presents this decomposition for the light neutrino-exchange mechanism, where one can see the cancellation between $I=0$ and $I=2$, 
similar to the case of $^{130}$Te and $^{136}$Xe  \cite{NeacsuHoroi2015}, $^{82}$Se \cite{SenkovHoroiBrown2014} and $^{48}$Ca  \cite{SenkovHoroi2013}. 
We perform this analysis in the case of heavy neutrino-exchange mechanism and we find a behavior similar to that of the light neutrino-exchange mechanism.
Fig. \ref{pair_heavy} displays the $I$-pair decomposition for the heavy neutrino-exchange mechanism.
Since the tensor contribution is small, we exclude it from the decomposition.
In Fig. \ref{pair_light} and Fig. \ref{pair_heavy} we use the effective values for the Fermi NME, by multiplying them with $-(g_V/g_A)^2$ (see Eq. (\ref{0nnme})). 
As a consequence, the bars in Fig. \ref{pair_light} and Fig. \ref{pair_heavy} can be added directly to get the total NME.
This representation is slightly different from the similar one of Refs. \cite{SenkovHoroiBrown2014,SenkovHoroi2014,NeacsuHoroi2015}, in which the raw sign of the Fermi contribution was considered. The new representation clearly shows that the shell model light neutrino-exchange NME are smaller due to the dramatic cancellation between the $I=0$ and $I=2$ contributions, while for the heavy neutrino-exchange NME the cancellation is not as pronounced (see also the discussion of the $^{76}$Ge case in Ref. \cite{BrownFangHoroi2015}).

The numerical results for the NME corresponding to the two mechanisms can be found in Tables \ref{tab_nme_comp_light} and \ref{tab_nme_comp_heavy} calculated for two SRC, Argonne-V18/Cd-Bonn.
If one uses a neutrino effective mass $m_{\beta\beta} \approx 0.1$ eV ($\mid \eta_{\nu L}\mid = 2\times 10^{-7}$), one obtains for $^{124}$Sn a $0\nu\beta\beta$ decay half-life in the range 
$2.3 \times 10^{26}-2.7 \times 10^{26}$ yr (the lower limit corresponds to CD-Bonn SRC). If the heavy neutrino mechanism dominates, for a  neutrino physics parameter $\mid \eta_{N R}\mid = 3\times 10^{-9}$ (the present experimental upper limit of this parameter is $7\times 10^{-9}$ \cite{Barry2013}) one gets a $0\nu\beta\beta$ decay half-life in the range $2.4 \times 10^{26}-5.0 \times 10^{26}$ yr. Here we used $G^{2\nu}_{01} = 9.06\times 10^{-15}$ yr$^{-1}$ \cite{Stefanik2015}. We also extracted the neutrino physics parameters, $\mid \eta_{\nu L}\mid \approx 10^{-6}$ (corresponding to $m_{\beta\beta} \approx 0.5$ eV) and $\mid \eta_{N R}\mid \approx 1.2\times 10^{-8}$ from the half-life lower limit for $^{136}$Xe using the NME of Ref. \cite{NeacsuHoroi2015}. Using these parameters we calculated the corresponding half-life lower limits for the transitions to the $0^+$ states of $^{124}$Te. The results are presented in Table \ref{tab_0nme}.

  \begin{figure}[htb] 
 \includegraphics[width=0.95\linewidth]{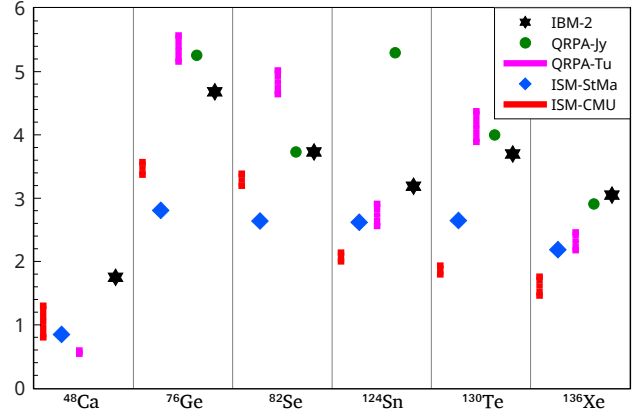}
 \caption{(Color online) Comparison of light neutrino-exchange $0\nu\beta\beta$ NME obtained with different nuclear structure methods. Order left to right in the data for each isotope corresponds to order down to up in the legend box.}
 \label{light}
 \end{figure}

  \begin{figure}[htb] 
 \includegraphics[width=0.95\linewidth]{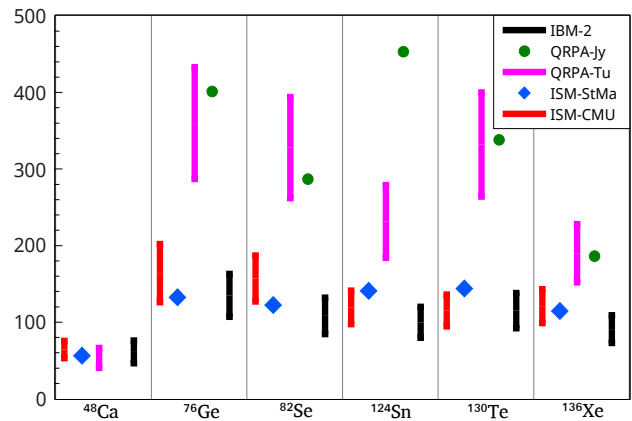}
\caption{(Color online) Same as Fig. \ref{light} for heavy neutrino-exchange NME.}
 \label{heavy}
 \end{figure}

\subsection{Overview of the $0\nu\beta\beta$ NME}\label{overview}

The NME are still responsible for the largest uncertainties in the half-life predictions, but recent theoretical efforts to restore the isospin symmetry in QRPA \cite{SimkovicRodin2013,Suhonen2015} and IBM-2 \cite{Barea2015} have significantly reduced some of the very large differences between NME results obtained with different nuclear structure methods. 
Given the recently reported QRPA \cite{Suhonen2015} and IBM-2 \cite{Barea2015} results, we decided to update the NME overview of Ref. \cite{NeacsuHoroi2015}, also adding the $^{124}$Sn $0\nu\beta\beta$ decay NME  for light and heavy neutrino-exchange mechanisms reported in this paper.  Figs. \ref{light} and  \ref{heavy} also show NME for the other five nuclei of experimental interest compared with results recently reported in the literature. In particular, we include the new heavy neutrino-exchange QRPA results from Jyvaskyla group \cite{Suhonen2015}.

Table \ref{tab_nme_comp_light} presents our new $^{124}$Sn light neutrino-exchange $0\nu\beta\beta$ decay NME results with two SRC parametrizations (Argonne-V18 and CD-Bonn), 
together with the NME of five other nuclei from our group (ISM-CMU), $^{48}$Ca \cite{SenkovHoroi2013}, $^{76}$Ge \cite{SenkovHoroi2014}, $^{82}$Se \cite{SenkovHoroiBrown2014}, $^{130}$Te \cite{NeacsuHoroi2015}, and $^{136}$Xe \cite{HoroiBrown2013,NeacsuHoroi2015}, 
compared with the most recent results that were obtained with nuclear structure methods that preserve the isospin symmetry and provide NME for both mechanisms. Included in Figure \ref{light} (but not in Table \ref{tab_nme_comp_light}) are also shell model light neutrino-exchange NME for $^{48}$Ca \cite{Kwiatkowsk2014}, $^{76}$Ge, and $^{82}$Se \cite{HoltEngel2013} that are using the same shell model calculations but an effective transition operator obtained in many-body perturbation theory. 
The updated IBM-2 NME \cite{Barea2015} are only available for the Argonne-V18 SRC. 
% Comparable QRPA UNC results are available only for $^{76}$Ge, $^{130}$Te and $^{136}$Xe from Ref. \cite{MustonendEngel2013}, using two Skyrme functionals (reflected in the error bar) and without SRC. 
The updated QRPA-Jy NME \cite{Suhonen2015} are only available for CD-Bonn SRC. 
The QRPA-Tu results \cite{FaesslerGonzales2014} are presented with both CD-Bonn and Argonne-V18 SRC. The bars in the figures correspond to the range of the two SRC (the higher end of the bars correspond to CD-Bonn SRC).
The ISM-StMa shell model NME results from Ref. \cite{MenendezPovesCaurier2009} are calculated with UCOM SRC and with effective Hamiltonians different of those used in the ISM-CMU calculations.

The new $^{124}$Sn heavy neutrino-exchange $0\nu\beta\beta$ decay NME results reported here are presented in Table \ref{tab_nme_comp_heavy}. 
Similar to the light neutrino-exchange case, we use two SRC parametrizations, and we include our results for the other five nuclei 
($^{48}$Ca \cite{SenkovHoroi2013}, $^{76}$Ge \cite{SenkovHoroi2014}, $^{82}$Se \cite{SenkovHoroiBrown2014}, $^{130}$Te \cite{NeacsuHoroi2015}, and $^{136}$Xe \cite{HoroiBrown2013,NeacsuHoroi2015}). They are compared with the results of the other methods included in Table \ref{tab_nme_comp_light}, when available.
 Since it does not appear to be much difference between the IBM-2 heavy neutrino NME reported in Ref. \cite{Barea2015} and those of Ref. \cite{Barea2013} we kept the old results, as they include  both CD-Bonn and Argonne-V18 SRC.
The QRPA-Jy results are from the recent Ref. \cite{Suhonen2015}, and the QRPA-Tu are from Ref. \cite{FaesslerGonzales2014}. The ISM-StMa are from Refs. \cite{Blennow2010,MenendezPrivate}.

Other methods investigating the light neutrino-exchange NME,
such as Projected Hartree Fock Bogoliubov \cite{Rath2013},
Energy Density Functional \cite{Rodriguez2010}, and the Relativistic
Energy Density Functional \cite{Song2014,Yao2015}, do not fall within our selection criteria, 
and are not included in our Figures.

\section{Conclusions} \label{conclusions}
 
This paper presents our shell model calculations for the double-beta decay of $^{124}$Sn in the $jj55$ model space, using the SVD effective Hamiltonian \cite{Chong2012} that was fine-tuned with experimental data for the Sn isotopes.
We validate the effective Hamiltonian by performing calculations of several spectroscopic quantities (energy spectra, $B(E2)\uparrow$ transitions),  finding good agreement with the experimental data. We also provide results for occupation probabilities and $GT$ strengths that could be investigated and validated by future experiments.

We provide half-lives for $2\nu\beta\beta$ decay to the ground state, $T^{1/2}_{2\nu}(g.s.) = 1.6 \times 10^{21}$ yr, 
 to the first excited $0^+_1$ state, $T^{1/2}_{2\nu}(0^+_1) = 6.2 \times 10^{26}$ yr, and to the first excited $2^+_1$ state, $T^{1/2(2^+_1)}_{2\nu}(2^+) = 2.2 \times 10^{26}$ yr, of $^{124}$Te. We estimate a theoretical uncertainty of at least $\pm 20\%$ for these predictions. 
The half-life for the g.s. is just beyond the current experimental limit \cite{Barabash2008}, but the transitions to the excited states are not accessible. If the $\left(0_1^+\right)^*$ state from Ref. \cite{Tableofisotopes8} is validated by future experimental investigations, then the half-life for the transition to this state is predicted to be $T^{1/2(0^+)}_{2\nu} = 5.6 \times 10^{23}$ yr. Our results are quite different from other similar results reported in the literature. We believe that our
 predicted $2\nu\beta\beta$ half-lives for $^{124}$Sn can be used to guide future experimental efforts, and potentially save considerable resources.

In the case of $0\nu\beta\beta$ decay, we report new $^{124}$Sn NME in the range $2.00-2.15$ for the light left-handed neutrino-exchange mechanism.
For the heavy right-handed neutrino-exchange mechanism, we obtain $^{124}$Sn NME in the range $97.4-141$.
We also present an analysis of the $I$-pair decomposition of the $^{124}$Sn NME for the light and heavy neutrino-exchange mechanisms. An optimal closure energy was used for the light neutrino-exchange NME.
In both cases we found that the main contribution to the NME is provided by the cancellation between 
$I=0$ and $I=2$ pairs, similar to the case of $^{48}$Ca \cite{SenkovHoroi2013}, $^{82}$Se \cite{SenkovHoroiBrown2014}, $^{130}$Te and $^{136}$Xe \cite{NeacsuHoroi2015}.
Using a neutrino effective mass  $m_{\beta\beta}\approx 0.1$ eV ($\mid \eta_{\nu L}\mid = 2\times 10^{-7}$), one obtains for $^{124}$Sn a $0\nu\beta\beta$ decay half-life lower limit 
$T_{0\nu}^{1/2} \approx 2.4 \times 10^{26}$ yr. Should the heavy neutrino mechanism dominate, this half-life would require a neutrino physics parameter $\mid \eta_{N R}\mid \approx 3\times 10^{-9}$ (the present experimental upper limit of this parameter is $7\times 10^{-9}$ \cite{Barry2013}).

Finally we present an up-to-date overview of the NME for both light neutrino and heavy neutrino-exchange mechanisms recently reported in the literature. We include in our comparison results of methods that enforce or restore the isospin symmetry, and that provide NME for both mechanisms. Although the approximate restoration of isospin symmetry lowers some of the highest NME, the ratio between the lowest and the highest NME for each isotope is still about 2 - 2.5 for the light neutrino-exchange mechanism, and about 3-5 for the heavy neutrino-exchange mechanism. One can also notice the reduced discrepancy between the IBM-2 and the shell model results.
The larger spread of heavy neutrino-exchange NME, compared to that of the light neutrino-exchange NME, also indicates a higher sensitivity of these calculations with respect to the SRC parameterization.
This uncertainty could be improved by obtaining an effective transition operator \cite{HoltEngel2013}, which is regularized together with the underlying nuclear Hamiltonian \cite{BognerHergert2014}.
 
\begin{acknowledgments}
%The authors thank B.A. Brown and R. Senkov for useful discussions and advise.
Support from  the NUCLEI SciDAC Collaboration under
U.S. Department of Energy Grant No. DE-SC0008529 is acknowledged.
MH also acknowledges U.S. NSF Grant No. PHY-1404442
\end{acknowledgments}

\bibliographystyle{apsrev}
\bibliography{dbd_bib}
 
\end{document}